\documentstyle[preprint,aps,multicol,graphicx,amsfonts]{revtex}

\graphicspath{{./}}

\begin{document}   
\newcommand{\bra}{\langle}
\newcommand{\ket}{\rangle}
\newcommand{\be}{\begin{equation}}
\newcommand{\ee}{\end{equation}}
\newcommand{\rd}{{\rm d}}
\newcommand{\br}{{\bbox{r}}}
\bibliographystyle{prsty}
\draft   
\widetext  
 \title{
An exact-diagonalization study of rare events in disordered conductors}
 \author{V. Uski$^1$, B. Mehlig$^2$, R. A. R\"omer$^1$ and M. Schreiber$^1$}
\address{
\mbox{}$^1$Institut f\"ur Physik, Technische Universit\"at Chemnitz,
           D-09107 Chemnitz, Germany\\
\mbox{}$^2$Theoretical Quantum Dynamics, Faculty of Physics, University of Freiburg,
          D-79104 Freiburg, Germany
        }
\date{\today}
\maketitle{ } 
\begin{abstract}  
We determine the statistical properties of wave functions
in disordered quantum systems by exact diagonalization
of one-, two- and quasi-one dimensional tight-binding
Hamiltonians. In the quasi-one dimensional case
we find that the tails of the distribution
of wave-function amplitudes 
are described by the non-linear $\sigma$-model.
In two dimensions, the tails of the distribution
function are consistent with a recent prediction
based on a direct optimal fluctuation method.
\end{abstract}   

It is well established
that disordered quantum systems
in the metallic regime (i.e., in the limit of weak disorder) and
highly excited classically chaotic quantum
systems exhibit universal quantum fluctuations
that can be described by random matrix theory (RMT):
statistical properties, on the scale of the mean level spacing, 
of eigenvalues, eigenfunctions, and matrix elements 
are  universal, i.e.,  they do not depend on
the microscopic details of the systems under
consideration \cite{meh67,por65,ber91,haa91,efe97}.

However, in ballistic, classically chaotic quantum
systems, non-hyperbolic phase-space
structures may lead to 
deviations from universal RMT statistics \cite{ber91}.
Similarly fluctuations in disordered, classically diffusive quantum
systems may deviate considerably from the RMT predictions 
due to increased localization. This 
effect is naturally very significant in the tails
of distribution functions  \cite{akl}
(corresponding to {\em rare events})
of wave-function amplitudes
\cite{alt89,mir93,fyo94,fal95,fyo95,mir97,mir99},
of the local density of states \cite{alt89,mir97},
of inverse participation ratios \cite{mir97,mir99}
and of NMR line shapes \cite{alt89}.
In all of these cases
(with the exception of Ref.~\cite{alt89} which deals
with one-dimensional (1D) systems), 
the distribution functions have been calculated
using the non-linear $\sigma$-model (NLSM).
Very recently, this approach has been
extended to ballistic systems \cite{muz95,aga95} (see also
\cite{bog96,pra96,meh99}).

In Ref.~\cite{smo97} 
a direct optimal fluctuation method \cite{zit66}
was used to calculate the tails of distributions
of current relaxation times and
wave-function amplitudes; and predictions
differing from \cite{mir93,fyo94,fal95,fyo95,mir97,mir99}
were put forward. This led the authors
of \cite{smo97} to question the suitability
of the NLSM to describe rare
events in disordered conductors.

It is thus of great
interest to test the predictions
of \cite{alt89,mir93,fyo94,fal95,fyo95,mir97,mir99}
and \cite{smo97} against results of independent calculations.
In this letter, we have determined distribution functions
of wave-function amplitudes by exact diagonalization
of 1D, 2D and quasi-1D 
tight-binding Hamiltonians; in this case rare events correspond
to unusually high splashes of wave-function amplitudes.
We note that wave-function amplitude
distributions can be measured in micro-wave
experiments \cite{srid,kuhl}.
 
We use the Anderson 
model of localization \cite{and58}  which is
a tight-binding model
on a $d$-dimensional hyper-cubic lattice  
\begin{equation}
\label{eq:defH}
\widehat{H} =  \sum_{\br,\br'} t^{\phantom \dagger}_{\br\br'} c_\br^\dagger \
c^{\phantom\dagger}_{\br'}
+ \sum_\br {\upsilon}^{\phantom\dagger}_\br c_\br^\dagger 
c^{\phantom\dagger}_\br\,.
\end{equation}
Here $\br = (x,y,\ldots)$ denotes sites on
the lattice, $c_\br^\dagger$ and $c_\br$ are the usual
creation and annihilation operators,
the hopping amplitudes are
$t_{\br\br'}=1$ for nearest neighbour sites
and zero otherwise.
The on-site potential $\upsilon_\br$ is 
taken to be uncorrelated white noise,
with  zero mean  and
variance $\langle \upsilon_\br\upsilon_{\br'}\rangle = \delta_{\br\br'} W^2/12$.
The parameter $W$ characterizes the disorder strength.
As is well-known (see for instance \cite{xxx,mul97}), 
the eigenvalues $E_j$ and
eigenfunctions  $\psi_j(\br)$ of this Hamiltonian,
in the metallic regime,
exhibit fluctuations described by RMT. 
In this case, Dyson's Gaussian orthogonal
ensemble \cite{meh67} is appropriate. 
When the matrix elements $t_{\br\br'}$ of (\ref{eq:defH})
are given an appropriate complex phase factor,
Dyson's unitary ensemble \cite{meh67} applies.
We refer
to these two cases by assigning, as usual,
the parameter $\beta=1$ to the former
and $\beta = 2$ to the latter. 
The metallic regime is characterized
by $g \gg1$ where $g = 2 \pi \nu V D L^{-2}$ 
is the dimensionless conductance (we take $\hbar=1$).
Here $\nu = 1/(V \Delta)$, $\Delta$ is the mean level spacing
and $V$ the volume.
$D= v_{\rm F}^2 \tau/d$ is the diffusion constant,
$\tau$ the mean free time and $v_{\rm F}$
the Fermi velocity. 
Four length scales are important: the lattice spacing $a$, the linear
extension $L$, the localization length $\xi$
and the mean free path $\ell = v_{\rm F}\tau$.

By diagonalizing the Hamiltonian $\widehat H$
using a  modified
Lanczos algorithm \cite{cul85},
we have determined the distribution function
\begin{equation}
\label{eq:defP}
f_\beta(E,\br;t) 
=\Delta\Big\langle\sum_j\!
\delta(t\!-\!|\psi_j(\bbox{r})|^2V)g_\eta(E\!-\!E_j)\Big  \rangle_W\,.
\end{equation}
Here
$\langle\cdots\rangle_W$ denotes an average over disorder realisations.
The wave functions are normalized
so that $\langle |\psi_j(\br)|^2\rangle_W = V^{-1}$
and $g_\eta(E)$ is a window function of width
$\eta$, centered around $E=0$ and normalized to unity.
In the following we describe the results of
our calculations and compare them to
the predictions of Refs.~\cite{mir93,fyo94,fal95,fyo95,mir97,mir99,smo97}.

\begin{figure}
 \centerline{\includegraphics[width=10.0cm]{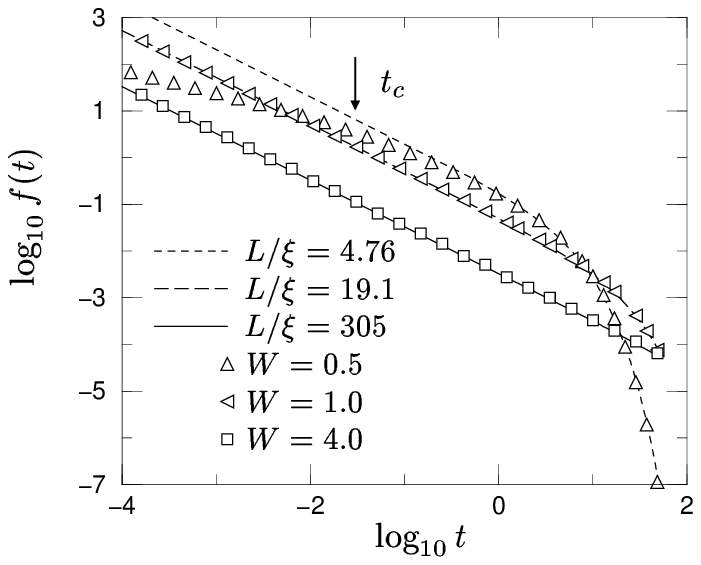}}
 \caption{\label{fig:1da}
 $\langle f(E,x;t)\rangle_x$ 
 for a chain  of length $L/a = 2000$ with periodic boundary conditions,
 $E=0$ and $\eta =  0.2$.  The lines
 are determined from Eq.~(\protect\ref{eq:1da}).
 The arrow indicates $t_{\rm c}$ for $L/\xi = 4.76$. }
 \end{figure}

{\em 1D case.}
The eigenstates in a  disordered chain 
are localized with
localization length $\xi = 4\ell$.
According to Ref.~\cite{alt89} the distribution
of wave-function amplitudes 
in a disordered chain of length $L$ is
\cite{note_xi}
\begin{equation}
\label{eq:1da}
f(E;t)\simeq\frac{\xi}{L t} \exp\Big(-\frac{ t\, \xi}{L}\Big)
\end{equation}
for $L\gg \ell$, independent of $x$ and $\beta$.
Our results for $\langle f(E,x;t)\rangle_x$
($\langle\cdots\rangle_x$ denotes an average
over the $x$ coordinate)
in Fig.~\ref{fig:1da} show very good agreement with Eq.~(\ref{eq:1da}) 
for large $L/\xi$.
The deviations at small $t$ for $L/\xi = 4.76$
are due to the fact that Eq.~(\ref{eq:1da})
is only valid asymptotically for large $L$.
It does not take into account 
that in a system of finite length $L$,
the smallest amplitude of a normalized,
exponentially decaying wave function
is of the order of $t_{\rm c} \simeq (L/\xi)\,\exp(-L/\xi)$.
This cut-off is shown in
Fig.~\ref{fig:1da}.

{\em Quasi-1D case}.
In  this case, as was shown
in Refs. \cite{mir93,fyo94},
the 
NLSM can be solved exactly 
for the distribution function $f_\beta(E,x;t)$,
using a transfer matrix approach \cite{efe83}.
The result is \cite{mir93,fyo94,mir97,mir99}
\be
\label{eq:q1do}
f_1(E,x;t)
= 
\frac{2\sqrt{2}}
{\pi \sqrt{t}}
\frac{\rd^2}{\rd t^2} \int_0^\infty\! \frac{\rd z}{\sqrt{z}}\,
Y(z+t/2) \,,
\ee
\be
\label{eq:q1du}
f_2(E,x;t)
= \frac{\rd^2}{\rd t^2} Y(t) \,.
\ee
Here 
$ Y(z) = {\cal W}(z \xi/L,x/\xi)\, {\cal W}(z \xi/L,(L-x)/\xi) $
and ${\cal W}(z,\tau)$ obeys the differential equation
\be
\frac{\partial }{\partial \tau} {\cal W}(z,\tau)
=\Big(z^2 \frac{\partial^2}{\partial z^2}-z\Big) {\cal W}(z,\tau)
\ee
with initial condition ${\cal W}(z,0) = 1$. The function
${\cal W}(z,t)$ may be determined in terms of 
an eigenfunction expansion of the operator $z^2 \partial_z^2-z$
\cite{fyo94,mir97,mir99}. The change of
the body {\em and} the tails of distribution $f_\beta(E,x;t)$ 
due to increasing localization 
may thus be parameterized by a single parameter 
which we define to be $X\equiv (\beta/2) L/\xi$.
Here $\xi  \equiv \beta \pi \nu D S$  where
$S$ is the cross-section of the wire.
Thus $X$ does not depend on $\beta$.
In the metallic regime (where $X \rightarrow 0$)
$Y(z) \simeq \exp(-z)$ which leads to the
usual RMT results $f^{(0)}_1(t) = \exp(-t/2)/\sqrt{2\pi t}$
and $f^{(0)}_2(t) = \exp(-t)$. The former distribution ($\beta = 1$)
is often referred to as the Porter-Thomas distribution
\cite{por65}. For increasing localization 
(finite but still small $X$), the ${\cal O}(X)$-corrections to
the body of the distribution function $f^{(0)}_\beta$
are obtained by expanding 
$Y \simeq \exp(-t)[1+\beta^{-1}t^2 P(x,x;0)]$
where $P(x,x';\omega)$ is the one-dimensional
diffusion propagator.
The result is $f_\beta(E,x;t) = f^{(0)}_\beta(t)[1+\delta f_\beta(E,x;t)]$
with
\begin{eqnarray}
\label{eq:q1ddev}
\delta f_\beta(E,x;t)  \!&\simeq &\!
P(x,x;0)\!\left\{
\begin{array}{ll}
3/4\!-\!3t/2\!+\!t^2/4 &\!\mbox{for$\,\beta\!=\!1$}\,,\\
1\!-\!2 t\!+\!t^2/2   &\!\mbox{for$\,\beta\!=\!2$}\,,
\end{array}
\right .
\nonumber
\end{eqnarray}
valid for $t \ll X^{-1/2}$.
In the tails ($t \gg X^{-1} > 1$) of $f_\beta(E,x;t)$,
Eqs.~(\ref{eq:q1do}) and (\ref{eq:q1du}) simplify to
\cite{fyo94}
\be
\label{eq:q1d-tails}
f_\beta(E,x;t) \simeq A_\beta(x,X)
\exp(-2\beta\sqrt{t/X})\,.
\ee
This result may also be obtained within
a saddle-point approximation to the NLSM \cite{fal95}.
The prefactors $A_\beta(x,X)$ for $\beta=1,2$ are
given in \cite{fyo94,fal95}.

According to Refs.~\cite{efe97,efe83} 
and \cite{mir93,fyo94,fal95,fyo95,mir97,mir99}
the 
NLSM applies provided the following 
conditions are satisfied:
\be
\label{eq:c1}
1 \ll k_{\rm F} \ell  \ll k_{\rm F}^2 S \ll k_{\rm F} L
\ee
where $k_{\rm F}$ is the Fermi wave vector
and $S$ is the cross section of the wire.
The first condition ensures that disorder is
sufficiently weak. The second condition
implies that apart from the sample
geometry, all other properties are
essentially 3D                \cite{efe83}.
Due to the third condition the return probability
is dominated by diffusive contributions \cite{mir99}. 
When  $k_{\rm F} = {\cal O}(a^{-1})$,  Eq.~(\ref{eq:c1}) corresponds to
$ 1 \ll \ell/a \ll M \ll L/a$,
where $M = k_{\rm F}^2 S$ is the number
of channels.
Furthermore, in the metallic regime, one must have $\ell \ll L \ll \xi$. 
Since $\xi \simeq M \ell$, this implies
\be
\label{eq:c2}
1 \ll L/\ell \ll M\,.
\ee
In a finite system,
the conditions  (\ref{eq:c1}) and (\ref{eq:c2})
are not easily met simultaneously.
We have performed exact diagonalizations
for $128\times 4 \times 4$  and
$128\times 8 \times 8$
lattices,
using open boundary conditions (BC) in the longitudinal
direction. 
In this case,
$P(x,x;0) = 2X\,[1/3-x(L\!-\!x)/L^2]$.

\begin{figure}
  \centerline{\includegraphics[width=10.0cm]{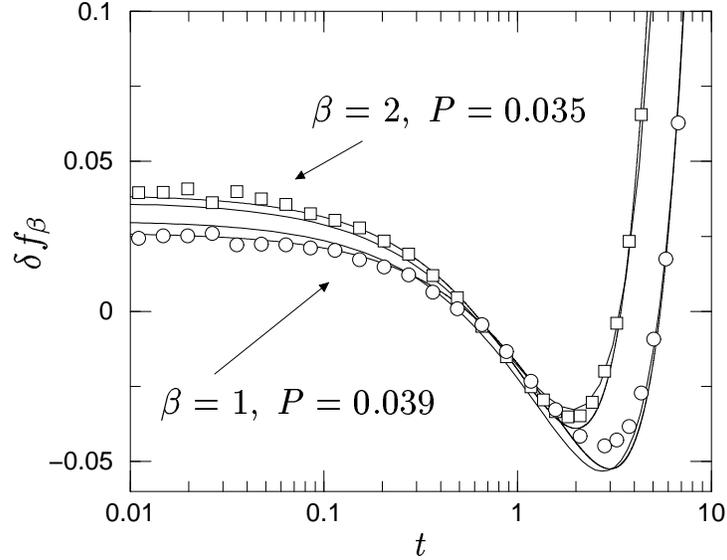}}
  \caption{\label{fig:q1ddev}
   $\langle \delta f_\beta(E,x;t)\rangle_x$
   for a $128\times 8\times 8$ lattice
   with $E=-1.7$, $\eta = 0.01$, and $W=1.0$.  For $\beta = 1$ 
   ($\protect\circ$) and $\beta=2$ ($\Box$), 
  these results are compared to Eqs.~(\ref{eq:q1do}),(\ref{eq:q1du})
   (\protect\rule[1.0mm]{1cm}{0.3mm}) and Eq.~(\ref{eq:q1ddev}) 
  (\protect\rule[1.0mm]{1cm}{0.1mm}). }
   \end{figure}

The results of our calculations are summarized
in Figs.~\ref{fig:q1ddev} and \ref{fig:q1d-loc}.
Figure~\ref{fig:q1ddev} shows  $\langle\delta f_\beta(E,x;t)\rangle_x$
in comparison with Eqs.~(\ref{eq:q1do}),(\ref{eq:q1du}) and (\ref{eq:q1ddev}).
We observe very good agreement.
The value of $P\equiv\langle P(x,x;0)\rangle_x$ should
be independent of $\beta$. 
As can be seen in Fig.~\ref{fig:q1ddev},
the value of $P$ does somewhat change with $\beta$,
albeit weakly.
For narrower wires ($128\times 4 \times 4$)
we have observed that the ratio
$P_1/P_2$ (determined by fitting $P\equiv P_\beta$
independently for $\beta=1,2$) becomes
very small for small values of $W$
(corresponding to $X \stackrel{<}{\sim} 0.1$)
while it approaches unity for large values of $W$.
A possible explanation for
this deviation would be that
for small $M$ {\em and} small $X$,
the condition (\ref{eq:c1})
is no longer satisfied
since $\ell/a \stackrel{>}{\sim} M $.
Surprisingly, the form of the
deviations is still very well described
by Eq.~(\ref{eq:q1ddev}) (not shown).

Figure~\ref{fig:q1d-loc} shows 
the tails of $f_\beta(E,x;t)$ for
weak disorder ($ X \stackrel{<}{\sim} 1$)
in comparison with Eqs.~(\ref{eq:q1do}),
(\ref{eq:q1du}) and (\ref{eq:q1d-tails}). 
Since for very small values of $X$
the tails decay so fast that
we cannot reliably calculate
them, we decreased the wire cross section
and increased the value of $W$
in Fig.~\ref{fig:q1d-loc}, thus increasing $X$.
The quoted values of $X$ were obtained
by fitting Eqs.~(\ref{eq:q1do}) and (\ref{eq:q1du}).
The values thus determined differ somewhat between $\beta=1$ and $2$
(see Fig.~\ref{fig:q1d-loc}) and this difference depends on
the choice of $E,\eta$ and $W$. 

In summary
we conclude that non-universal
deviations from RMT statistics in
quasi-1D wires
are very well described
by a NLSM not only
in the {\em body} (see Fig.~\ref{fig:q1ddev})
but notably also in the {\em tails} (see Fig.~\ref{fig:q1d-loc})
of the distribution
$f_\beta(E,x;t)$.

{\em 2D              case}. 
In this case, according to Ref.~\cite{fyo95},
corrections to $f^{(0)}_\beta$
are still given
by Eq.~(\ref{eq:q1ddev}), but
now $P=\langle P(\br,\br;0)\rangle_\br$ where
$P(\br,\br';\omega)$ is the 2D diffusion
propagator.
For the tails of the distributions, the result
of the NLSM is within
a saddle-point approximation \cite{fal95,mir99}
\be
\label{eq:nls_tails}
f_\beta(E,\br;t) \simeq 
\exp\big[-C_\beta (\mbox{ln} t)^2 \big]
\ee 
with 
\be
\label{eq:Cnls}
C_\beta = \beta\pi g/[4\,\mbox{ln}(L/\ell)]\,.
\ee
Note that according to (\ref{eq:Cnls})
the decay in the  tails of Eq.~(\ref{eq:nls_tails})
depends on $\beta$, as in the quasi-1D              case
[Eq.~(\ref{eq:q1d-tails})].

Recently, in Ref.~\cite{smo97} a 
different approach (a direct optimal
fluctuation method \cite{zit66})
was used to calculate the tails of $f_\beta(E,\br;t)$.
According to Ref.~\cite{smo97}, the tails
of the distribution function are given by Eq.~(\ref{eq:nls_tails})
but with $C_\beta$ replaced by
\be
\label{eq:langer_tails}
C = \pi g/[2\,\mbox{ln}(L/r_0)]
\ee 
[with $r_0 = {\cal O}(k_{\rm F}^{-1})$]
which differs in two respects from the
prediction of the NLSM:
First, $\ell$ in $C_\beta$
is replaced by $r_0$ 
in (\ref{eq:langer_tails}).
Second, 
there is no $\beta$-dependence.

We have diagonalized the Hamiltonian
(\ref{eq:defH}) on a $100\times 100$ lattice.
Figure~\ref{fig:2ddev} shows corrections to 
$f_\beta^{(0)}$ for weak disorder.
We find that the form of the deviations is
very well described by Eq.~(\ref{eq:q1ddev}). However,
the values of $P_\beta$ obtained for $\beta=1,2$
differ by a factor $<1/2$.
  A possible
  explanation for this deviation
  might be the following \cite{mir99}:
  In the {\em ballistic} regime, $P$
  is no longer given by the diffusion
  propagator but may be dominated by a single-scattering
  expression which involves an additional
  factor $\beta/2$ and thus
  $P_1/P_2 <1$.  
  It would be tempting to deduce
  from this that in our case
  {\em ballistic} effects are important.
  However, this does not explain
  $P_1/P_2 < 1/2$. 
  Numerical results \cite{usk99}
 (albeit for rather small systems) indicate
 that in 3D, $P_1/P_2 \simeq 1/2$
 for the parameters chosen in \cite{usk99}.

Figure~\ref{fig:2dtail} shows the tails
of the distribution functions. 
The tails are consistent with an $\exp[-C(\ln t)^2]$
decay as predicted by Eq.~(\ref{eq:nls_tails}).
We have thus verified 
that corrections to RMT distributions in
2D              systems do 
give rise to log-normal tails.
Our results suggest that the prefactor
in the exponent does not depend on $\beta$.
 This result  is  consistent with
 Eq.~(\ref{eq:langer_tails}).
We have independently calculated
the dimensionless conductance $g$ 
using the usual linear response expression.
We find that $g$ is independent
of $\beta$ and $g \propto W^{-2}$,
as expected (inset of Fig.~\ref{fig:2dtail}).

The inset of Fig.~\ref{fig:2dtail} shows
that $C$ increases with  decreasing disorder
strength, as it should, albeit slower
than $W^{-2}$. The increase of $C$ for
decreasing $W$ is underestimated, because for
weak disorder, the tails of the distributions
have not reached the asymptotic regime.

In summary we have reported on
a study of rare events
in disordered conductors,
by diagonalizing
the tight-binding Hamiltonian (\ref{eq:defH})
and analyzing the probability
of rare splashes of high wave-function amplitudes. 
Our 1D results agree with those of \cite{alt89}. 
In the quasi-1D case, 
we have compared our
data to an exact solution \cite{mir93,fyo94} of the 
NLSM, and to a saddle-point approximation \cite{fal95}.
We observe very good agreement
between our results and
those of the calculations based
on the NLSM and thus conclude
that the NLSM provides a quantitative
description of rare events in quasi-1D                 
disordered conductors.
In 2D systems, corrections
to the body of the distribution
functions  are well described
by results based on
the NLSM, with a modified prefactor $P_\beta$.
Moreover, we could verify that the tails of
the distribution function in the vicinity
of the metallic regime are log-normal.
Thus our numerical investigations, which are 
complementary to the analytical predictions, corroborate the overall 
picture suggested in Refs.~\cite{mir93,fyo94,fal95,fyo95,mir97,mir99}
for quasi-1D and 2D systems. The coefficient describing the tails of
wave-function distributions in 2D systems turns out to be 
independent of $\beta$ for the parameters considered in this paper, as opposed
to the quasi-1D case. This is
consistent with the prediction of the direct
optimal fluctuation method (Ref.~\cite{smo97}).

{\em Acknowledgment}. This work was supported by
the DFG under project C5/SFB393.

\newpage
  {\begin{figure}
  \centerline{\includegraphics[width=10.0cm]{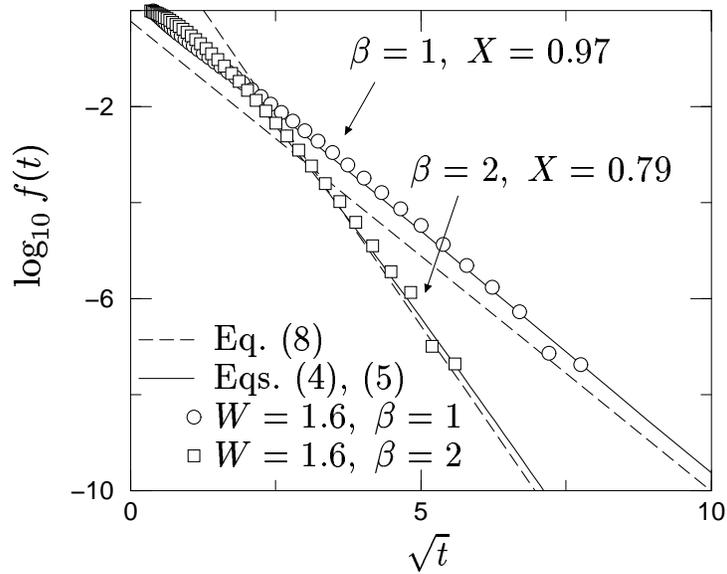}}
  \caption{\label{fig:q1d-loc}
  $f_\beta(E,x;t)$ for a $128\times 4\times 4 $
   lattice; for $X \stackrel{<}{\sim} 1$, 
   $x \simeq L/2$, $W=1.6$, $E=-1.7$, $\eta=0.01$ 
  and $\beta=1,2$ compared to Eqs.~(\ref{eq:q1do}), (\ref{eq:q1du})
  and (\ref{eq:q1d-tails}).  }
  \end{figure}

  \begin{figure}
  \centerline{\includegraphics[width=10.0cm]{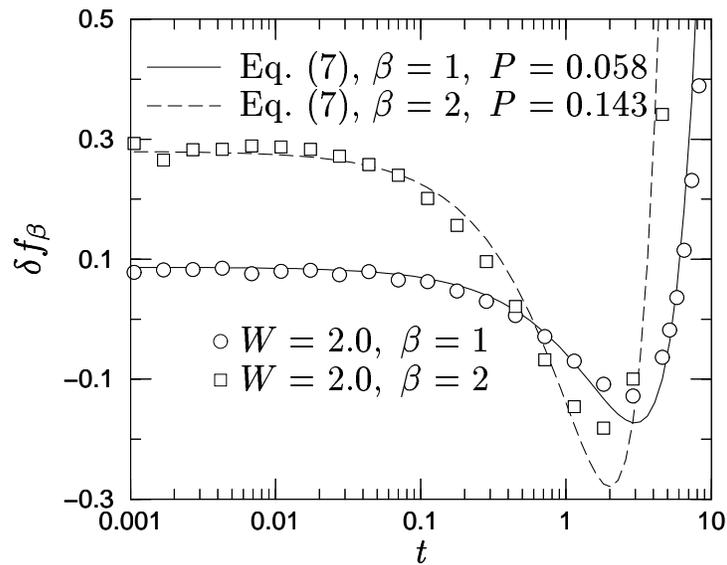}}
  \caption{\label{fig:2ddev}
  $\langle \delta f_\beta(E,\br;t)\rangle_{\br}$
  for a
  $100\times 100$ lattice with periodic BC, 
  for  $E=-1.6$ and $\eta = 0.005$.
 }
  \end{figure}
\newpage
  \begin{figure}
  \centerline{\includegraphics[width=10.0cm]{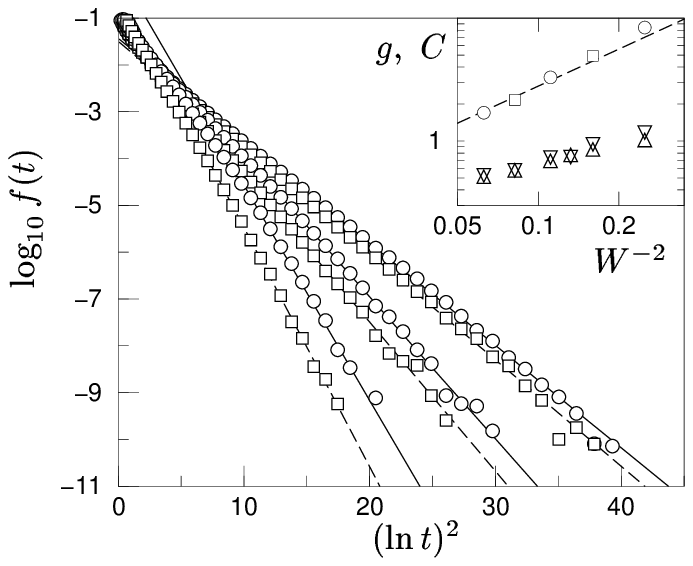}}
  \caption{\label{fig:2dtail}
  Tails of $\langle f_\beta(E,\br;t)\rangle_{\protect\br}$ 
  for the same system as in Fig.~\protect\ref{fig:2ddev}, 
 for $W = 2,3$ and $4$, for
  $\beta = 1$ ($\protect\circ$) and $\beta = 2$ ($\protect\Box$).
   The lines show Eq.~(11). The inset shows the fitted values of $C$
  versus $W^{-2}$, for $\beta=1$ ($\bigtriangleup$)
  and  $2$ ($\bigtriangledown$)
  and the dimensionless conductance $g$ for
  $\beta=1$ ($\circ$) and $2$ ($\Box$). 
  The dashed line indicates $W^{-2}$ behaviour.}
  \end{figure}


\begin{thebibliography}{10}
\bibitem{note_xi} Here $\xi$ is defined so that a typical wavefunction
               decays according to $|\psi(x)| \sim \exp(-|x-x_0|/\xi$) away
               from the localization center $x_0$. 
               With this choice of $\xi$, one has $\xi = 105/W^2$
               to second order in $W$.
\end{thebibliography}
\end{document}